\documentclass[journal]{IEEEtran}

\usepackage{amsmath}
\usepackage{amsfonts}
\usepackage{amssymb}
\usepackage{epsf}
\usepackage{cite}
\usepackage{balance}
\usepackage{fancyhdr}
\usepackage{graphicx}
\usepackage{subfigure}
\usepackage[blocks]{authblk}
\usepackage[stable]{footmisc}
\usepackage{float}
\usepackage{fancyhdr}
\usepackage{gensymb}
\usepackage[bookmarks=false]{hyperref}
\usepackage{epstopdf}

\makeatletter
\def\ps@IEEEtitlepagestyle{%
  \def\@oddfoot{\mycopyrightnotice}%
  \def\@evenfoot{}%
}
\def\mycopyrightnotice{%
  {\footnotesize \copyright 2015 IEEE. Personal use of this material is permitted. Permission from IEEE must be obtained for all other uses, in any current or future media\hfill}% <--- Change here
  \gdef\mycopyrightnotice{}% just in case
}

\newcounter{subeq}

\IEEEoverridecommandlockouts
\date{}
\begin{document}
\title{Impact of Multipath Reflections on the Performance of Indoor Visible Light Positioning Systems}

\author{Wenjun Gu,~\IEEEmembership{Student Member,~IEEE,}
        Mohammadreza~Aminikashani,~\IEEEmembership{Student Member,~IEEE,}
        and~Mohsen~Kavehrad,~\IEEEmembership{Fellow,~IEEE}

\thanks{Wenjun Gu, Mohammadreza Aminikashani and Mohsen Kavehrad are with the Pennsylvania State University, University Park, PA 16802 (email:  {wzg112, mza159, mkavehrad}@psu.edu).}}

\maketitle
%\balance
\begin{abstract}
Visible light communication (VLC) using light-emitting-diodes (LEDs) has been a popular research area recently. VLC can provide a practical solution for indoor positioning. In this paper, the impact of multipath reflections on indoor VLC positioning is investigated, considering a complex indoor environment with walls, floor and ceiling. For the proposed positioning system, an LED bulb is the transmitter and a photo-diode (PD) is the receiver to detect received signal strength (RSS) information. Combined deterministic and modified Monte Carlo (CDMMC) method is applied to compute the impulse response of the optical channel. Since power attenuation is applied to calculate the distance between the transmitter and receiver, the received power from each reflection order is analyzed. Finally, the positioning errors are estimated for all the locations over the room and compared with the previous works where no reflections considered. Three calibration approaches are proposed to decrease the effect of multipath reflections.
 \end{abstract}
\begin{IEEEkeywords}
Indoor positioning, visible light communication, multipath reflections, impulse response, received signal strength.
\end{IEEEkeywords}

\section{Introduction}\label{INTRODUCTION}
\IEEEPARstart{L}{Location based services} (LBS) have become a popular research topic for several years which provide users with current locations and related services. For outdoor environment, Global Positioning System (GPS) provides satisfactory services such as localization, navigation and displaying surrounding traffic conditions. For indoor environments, GPS technology is not applicable since a satellite signal suffers from severe attenuation when passing though solid walls. In recent decades, several methods have been proposed to realize indoor positioning with the help of technologies such as ultra-wide band (UWB), wireless local area network (WLAN), Radio-frequency identification (RFID), Bluetooth and cellular system \cite{1}.

Light-emitting-diode (LED) technology has been developing very rapidly in recent decades. It not only provides people with economical and efficient illumination and a long service time, but also paves the way for smart lighting and visible light communication (VLC) \cite{2,peng,3, reza,3.1}. As a strong candidate for high-speed wireless networks of the next generation, visible light communication exhibits many advantages over conventional radio frequency (RF) communication. First, visible light, together with infrared and ultraviolet spectral band, provides unregulated and unlimited bandwidth as a practical solution to the current spectrum crunch issue \cite{4,ofdm}. Second, considering that light waves are unable to penetrate through solid walls and are confined in an individual room, the band reuse among different rooms is accomplished and physical layer security for the communication system is guaranteed. Third, VLC can be widely applied in many RF sensitive environments such as mines, power plants and hospitals due to the fact that light waves never generate any electromagnetic interference. Fourth, so long as illumination infrastructure exists, VLC is applicable so that the hardware cost is decreased.

The application of VLC technology for the indoor positioning has been extensively studied as an available solution for the LBS  \cite{4.1,7,6,5,8,ofdmindoor,9,10,11}. LED light sources act as transmitter and receiver is a photo-diode (PD) or an image sensor collocated with a user. Several approaches have been proposed to realize visible light positioning. In one approach, an image sensor is used to obtain angulation algorithm to calculate the receiver position based on angle-of-arrival (AOA) information and rotation matrix \cite{7}. In this method, colored LEDs are used to help the image sensor to distinguish between different light sources. Scene analysis is another approach to obtain the receiver position. Features of each location are collected as the fingerprints in the offline stage. In the online stage, the features of current location are measured and by matching those with offline fingerprints, location of receiver is estimated \cite{6}. In this paper, we employ a commonly used algorithm where the RSS information is first detected by a PD, and then distance between transmitter and receiver is calculated. The lateration algorithm is finally applied to estimate receiver coordinates \cite{5}.

In addition to the above methods, other technologies are introduced in VLC system to improve the positioning performance. Zigbee technology can be combined with VLC to realize long distance positioning \cite{8}. In \cite{9}, with the assistance of a 6-axes sensor (geomagnetic sensor and gravity acceleration sensor), a switching estimated receiver positioning system is proposed to achieve higher accuracy. Hybridizing accelerometer was proposed in \cite{10} to realize three dimensional positioning without knowledge of receiver height. Gaussian mixture sigma point particle filter can be further employed to achieve high positioning accuracy and prevention of large deviations \cite{11}.

In the literature, line-of-sight (LOS) channels have been considered without taking account of multipath reflections in analysis of positioning performance. However, transmitted signal introduces multipath reflections as it bounces off walls, ceiling and floor where the transmitter is a wide-beam LED source, and the receiver having a finite field-of-view (FOV) captures reflected photons from room surfaces. In this paper, the effect of multipath-induced distortion on positioning accuracy of indoor VLC positioning systems is investigated. Several methods have been proposed to approximate impulse response of an indoor optical wireless channel. In \cite{12}, Barry et al. proposed a deterministic algorithm that partitions a room into many elementary reflectors and sums up the impulse response contributions from different orders of bounces. As this method is recursive, it takes a significant amount of computation time. Monte Carlo ray tracing approach is an alternative way to calculate the impulse response where rays with identical optical power are traced from the source \cite{13}. The direction of rays are generated by a probability density function (PDF) modeled using a Lambertian pattern. When the rays hit reflecting surfaces, new rays with reduced power are generated from the impact point with the same PDF. This method suffers from the fact that it needs a very large number of rays while only a small portion of rays will finally reach the PD. To alleviate this issue, modified Monte Carlo (MMC) approach is proposed which exploits each ray several times instead of only once \cite{14}. Although this method is fast, it introduces some variance due to random direction of the rays. In \cite{21}, Alqudah and Kavehrad proposed a new approach to characterize diffuse links in a multiple-input multiple-output (MIMO) system. In this paper, we use the methodology developed recently in \cite{15}, referred as combined deterministic and modified Monte Carlo (CDMMC), taking advantage of both methods to simulate impulse responses of indoor optical wireless channels.

The rest of the paper is organized into five sections. In Section II, the system model and CDMMC approach used to calculate the impulse response are briefly discussed. In Section III, the positioning algorithm is described to estimate receiver coordinates. In Section IV, the effect of multipath reflections on the positioning system is thoroughly investigated. In Section V, three calibration methods are applied to increase the positioning accuracy. Section VI finally concludes the paper.

\section{Multipath Analysis}
\subsection{System Design}
\begin{figure}
\centering
\includegraphics[width = 8cm, height =6cm]{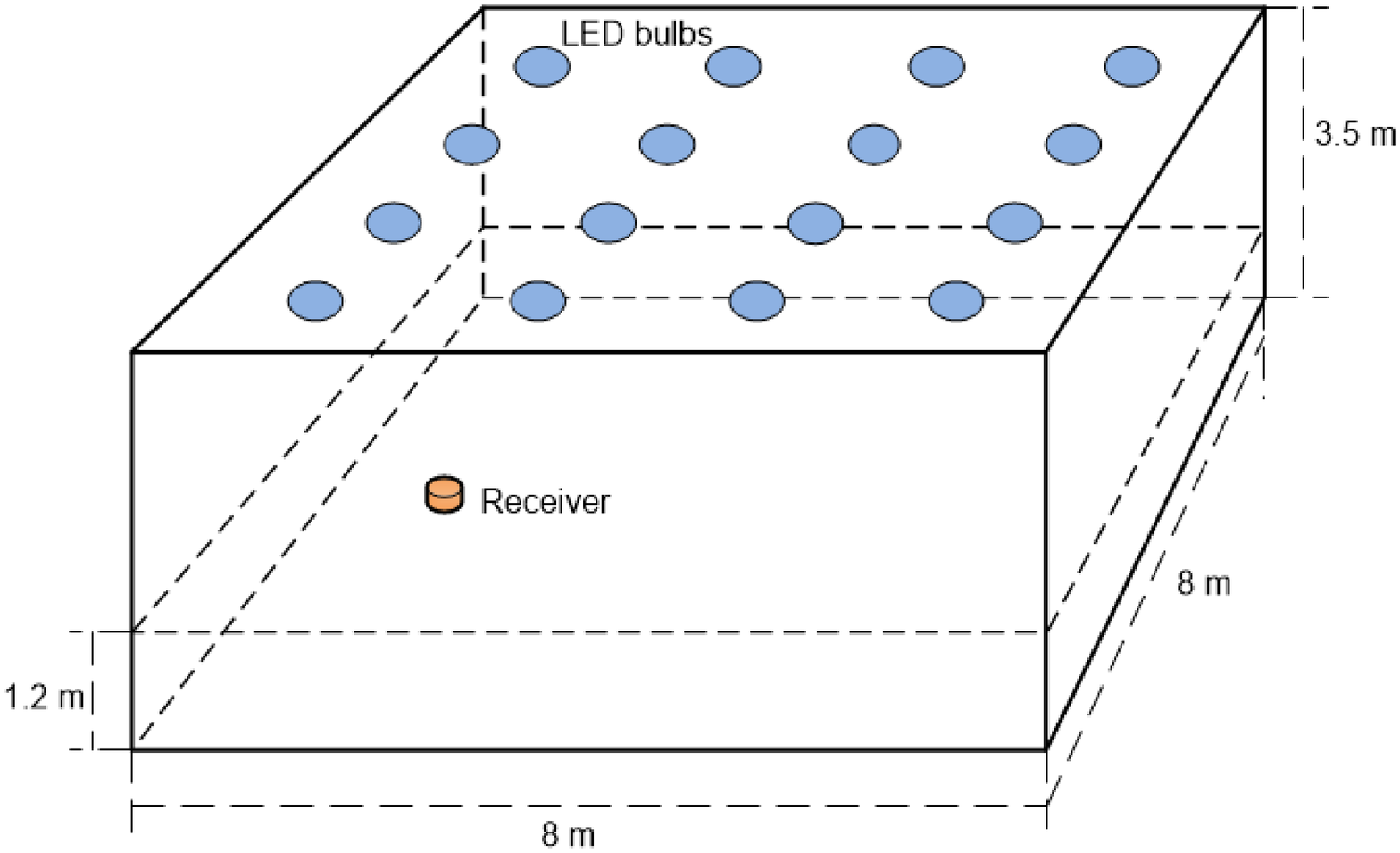}
\caption{System Configuration.}
\end{figure}
A typical indoor optical positioning system is shown in Fig. 1 where the LED bulbs as transmitters are installed on the ceiling of the room, and the receiver is located 1.2 m from the floor. Each of the transmitters has an identification (ID) code denoting its coordinates. On-Off-keying (OOK) modulation is used to modulate the LED bulbs. Six reflection surfaces of the room, i.e., four walls, one ceiling and one floor are assumed to be perpendicular with one another.

As shown in Table 1, the reflection coefficients are assumed to be fixed considering the material of the room surface and the 420 nm light source. The transmitters in the algorithm are treated as point sources and located at the height of 3.3 m considering practical installations. The distance between the transmitters is 2 m. As the transmitters are facing downwards, the azimuth angle is 0$^{\circ}$ and the elevation angle is -90$^{\circ}$. The receiver is facing upwards, thus the azimuth angle is 0$^{\circ}$ and the elevation angle is +90$^{\circ}$. The receiving area of the PD is 1$\times 10^{-4} \text{m}^{2}$, with 70$^{\circ}$  FOV. These parameters are summarized in Table 1.
\begin{table}
\begin{center}
\begin{quote}
\caption{Room Configuration}
\label{table1}
\end{quote}
%\resizebox{\columnwidth}{!}{
\begin{tabular}{|c|c|}
  \hline
  \textbf{Room dimensions} &\textbf{Reflection coefficients} \\ \hline
  length: 6 m &    ${{\rho }_{wall}}$: 0.66\\
  width: 6 m  & ${{\rho }_{Ceiling}}$: 0.35\\
  height: 3.5 m & ${{\rho }_{Floor}}$: 0.60\\ \hline
  \multicolumn{2}{|c|}{Horizontal coordinates of LED bulbs:}\\ \hline
  \multicolumn{2}{|c|}{(1.0 , 1.0) (1.0 , 3.0) (1.0 , 5.0) (1.0 , 7.0)}\\
  \multicolumn{2}{|c|}{(3.0 , 1.0) (3.0 , 3.0) (3.0 , 5.0) (3.0 , 7.0)}\\
  \multicolumn{2}{|c|}{(5.0 , 1.0) (5.0 , 3.0) (5.0 , 5.0) (5.0 , 7.0)}\\
  \multicolumn{2}{|c|}{(7.0 , 1.0) (7.0 , 3.0) (7.0 , 5.0) (7.0 , 7.0)}\\ \hline
  \end{tabular}
\end{center}
\end{table}
\subsection{Channel Access Method}
As the LED bulbs transmit their coordinates information independently, the signals cannot be retrieved if they interfere with each other at the receiver. To solve this problem, time division multiplexing (TDM) is applied as the channel access method \cite{16}. All the transmitters have synchronized frames and occupy different time slots in one frame period to send their signals. The frame structure is presented in Fig. 2. When one LED bulb transmits the ID information, other LED bulbs emit constant light intensity for the illumination purpose.
\begin{figure}
\centering
\includegraphics[width = 8cm, height = 4.5cm]{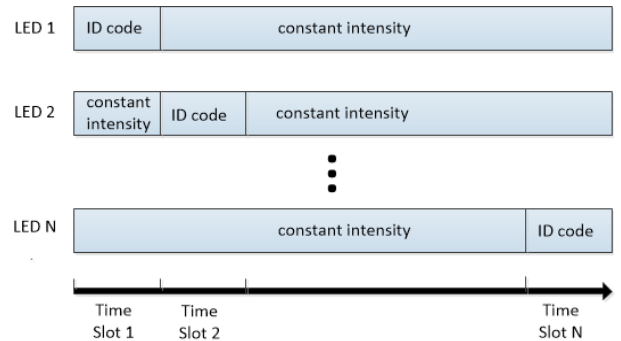}
\caption{Frame structure of the positioning system for one period.}
\end{figure}
\subsection{Impulse Response Analysis Method}
The CDMMC method combines the deterministic and the MMC methods to take advantage of both. The deterministic method is employed to calculate the contribution of the first reflections for high accuracy \cite{12}, while the second and higher-order reflections are calculated by the MMC method to achieve a high computational speed \cite{13}.

The first step is to divide the room surfaces into many small square elements each of which has an area of 1$\times 10^{-4} \text{m}^{2}$, equal to the PD’s receiving area.

Second, the PD together with these elements act as the receivers. The received power is obtained as
 \begin{equation}\label{eq1}
 {{P}_\text{received}}^{\left( 0 \right)}=H\left( 0 \right){{P}_\text{source}}^{\left( 0 \right)},
 \end{equation}
where ${{P}_\text{source}}^{\left( 0 \right)}$ is the power emitted from each LED bulb. In (\ref{eq1}), $H\left( 0 \right)$ is the channel DC gain \cite{17}, and ${{P}_\text{source}}^{\left( 0 \right)}$ is the received power for the LOS link.

Third, each of these small elements is considered as a point source whose power is calculated as
\begin{equation}\label{eq2}
{{P}_\text{source}}^{\left( 1 \right)}={{P}_\text{received}}^{\left( 0 \right)}{{\rho }_\text{surface}},
\end{equation}
where ${{\rho }_\text{surface}}$ is the reflection coefficient of ceiling, floor or walls listed in Table I.

In the fourth step, the received power of the first reflections is calculated by considering the small elements and PD as the receivers again which can be expressed as
\begin{equation}\label{eq3}
{{P}_\text{received}}^{\left( 1 \right)}=H\left( 0 \right){{P}_\text{source}}^{\left( 1 \right)}.
\end{equation}
The MMC method is then employed such that 10 random rays are generated from each small element where the PDF of the rays' directions follows
\begin{equation}\label{eq4}
{f\left( \alpha, \beta\right)=\frac{m+1}{2\pi }{{\cos }^{m}}\left( \alpha  \right)}.
\end{equation}
In (\ref{eq4}), $\alpha$ is the angle between $z$-axis and the ray vector shown in Fig. 3, $\beta$ is the angle between projection of the ray vector on the $X-Y$ plane and $x$-axis, and $m$ is the Lambertion order. In Fig. 3, the origin point is each ray's point source and the $X-Y$ plane represents the surface plane of the source. Note that (\ref{eq4}) is independent of $\beta$ .These rays arrive at the surface of the room and then the received power of secondary reflections is calculated as
\begin{equation}\label{eq5}
{{P}_\text{source}}^{\left( 2 \right)}={{{P}_\text{received}}^{\left( 1 \right)}{{\rho }_\text{surface}}}/{10}.
\end{equation}
\begin{figure}
\centering
\includegraphics[width = 7cm, height = 6cm]{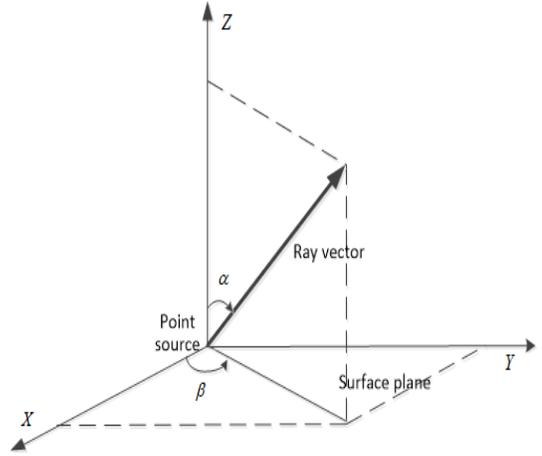}
\caption{Coordinate system for a random ray.}
\end{figure}
The impact points of these rays on the room surfaces are considered as the new sources, and then new rays are generated to calculate the subsequent reflections, recursively. Fig. 4 shows the flow diagram of the CDMMC method. The impulse response of the channel is computed by adding up all the contributions from LOS and those of reflections.
\begin{figure}
\centering
\includegraphics[width = 7cm, height =9cm]{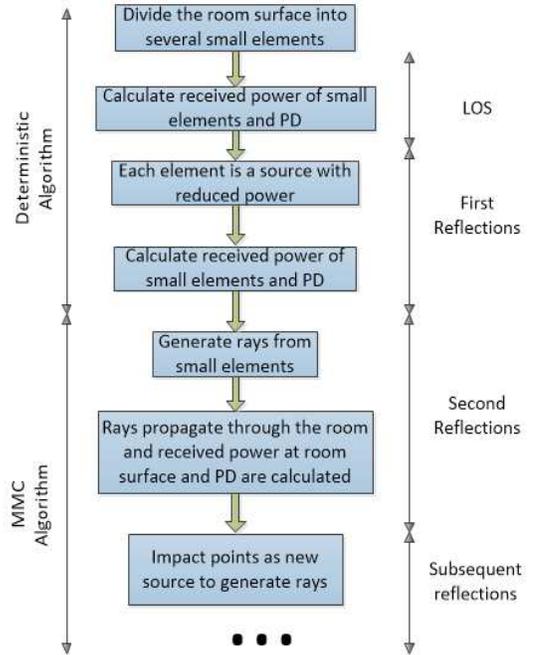}
\caption{Flow diagram of CDMMC method.}
\end{figure}
\section{POSITIONING ALGORITHM}
\subsection{Distance Estimation}
 After the impulse response is computed, received signal strength for transmitter $i$ is obtained as $P_{r}^{\left( i \right)}$, we assume that the transmitted power, i.e. ${{P}_{t}}$, is 5 W for logic “1” and 3 W for logic “0”. The relation between $P_{r}^{\left( i \right)}$ and ${{P}_{t}}$ can be expressed as \cite{17}
\begin{equation}\label{eq6}
P_{r}^{\left( i \right)}=\frac{m+1}{2\pi {{d}_{i}}^{2}}A{{\cos }^{m}}\left( \phi  \right){{T}_{s}}\left( \psi  \right)g\left( \psi  \right)\cos \left( \psi  \right){{P}_{t}},
\end{equation}
where $\phi$ is the irradiance angle from the transmitter to the receiver, $\psi$ is the incident angel, and ${{T}_{s}}\left( \psi  \right)$ is the transmittance function of the optical filter. In (\ref{eq6}),  $g(\psi )$ is the compound parabolic concentrator gain of the receiver  \cite{17}, and $A$ is the detector physical area. Table II presents the numerical value of these parameters assumed throughout this paper.
Considering transmitters and receiver's orientation, $\cos \left( \phi  \right)=\cos \left( \psi  \right)=h/{{d}_{i}}$ where $h$ is the vertical distance between transmitter and receiver. Therefore, the distance ${{d}_{i}}$ between the transmitter $i$ and the receiver is calculated as
\begin{equation}\label{eq7}
{{d}_{i}}=\sqrt[4]{\frac{\left( m+1 \right)A{{T}_{s}}\left( \psi  \right)g\left( \psi  \right){{P}_{t}}{{h}^{2}}}{2\pi P_{r}^{\left( i \right)}}}.
\end{equation}
\begin{table}
\begin{center}
\begin{quote}
\caption{Parameters of the transmitter and receiver}
\label{table2}
\end{quote}
%\resizebox{\columnwidth}{!}{
\begin{tabular}{|c|c|}
  \hline
  \textbf{Transmitters (Sources)} &\textbf{Receiver} \\ \hline
  Wavelength: 420 nm &Area $\left(A\right)$: 1$\times 10^{-4} \text{m}^{2}$\\
  Height $\left(H\right)$: 3.3 m &Height $\left(h\right)$: 1.2 m\\
  Lambertian mode $\left(m\right)$: 1 &Elevation: +90$^{\circ}$\\
  Elevation: -90$^{\circ}$ &Azimuth: 0$^{\circ}$\\
  Azimuth: 0$^{\circ}$ &FOV $\left({{\Psi }_{c}}\right)$: 70$^{\circ}$\\
   &${{T}_{s}}\left( \psi  \right)$:1 \\
   &$g\left( \psi  \right)$: 2.5481\\ \hline
\end{tabular}
\end{center}
\end{table}
The horizontal distance is finally estimated as
\begin{equation}\label{eq8}
{{r}_{i}}=\sqrt{d_{i}^{2}-{{h}^{2}}}=\sqrt{\sqrt{\frac{\left( m+1 \right)A{{T}_{s}}\left( \psi  \right)g\left( \psi  \right){{P}_{t}}{{h}^{2}}}{2\pi P_{r}^{\left( i \right)}}}-{{h}^{2}}}.
\end{equation}
\subsection{Least Square Estimation}
The receiver coordinates $\left( x,y \right)$ correspond to the transmitter coordinates $\left( {{x}_{i}},{{y}_{i}} \right)$ based on the following group of equations:
\begin{equation}\label{eq9}
 \begin{cases}
  {{\left( x-{{x}_{1}} \right)}^{2}}+{{\left( y-{{y}_{1}} \right)}^{2}}=r_{1}^{2}& \\
  {{\left( x-{{x}_{2}} \right)}^{2}}+{{\left( y-{{y}_{2}} \right)}^{2}}=r_{2}^{2}& \\
  \vdots & \\
  {{\left( x-{{x}_{n}} \right)}^{2}}+{{\left( y-{{y}_{n}} \right)}^{2}}=r_{n}^{2}& \\
\end{cases},
\end{equation}
where $n$ is the number of transmitters whose signal can be detected by the receiver. By subtracting the last three equations from the first one, we obtain
\begin{equation}\label{eq10}
\begin{cases}
\left( {{x}_{1}}-{{x}_{2}} \right)x+\left( {{y}_{1}}-{{y}_{2}} \right)y&\\
=\left( r_{2}^{2}-r_{1}^{2}-x_{2}^{2}+x_{1}^{2}-y_{2}^{2}+y_{1}^{2} \right)/2 &\\
\vdots &\\
\left( {{x}_{1}}-{{x}_{n}} \right)x+\left( {{y}_{1}}-{{y}_{n}} \right)y&\\
=\left( r_{n}^{2}-r_{1}^{2}-x_{n}^{2}+x_{1}^{2}-y_{n}^{2}+y_{1}^{2} \right)/2 &\\
\end{cases},
\end{equation}
which can be expressed in the matrix format as
\begin{equation}\label{eq11}
\mathbf{AX}=\mathbf{B}
\end{equation}
where
\begin{equation}\label{eq12}
\mathbf{A}=\left[ \begin{matrix}
   {{x}_{2}}-{{x}_{1}} & {{y}_{2}}-{{y}_{1}}  \\
   \vdots & \vdots  \\
   {{x}_{n}}-{{x}_{1}} & {{y}_{n}}-{{y}_{1}}  \\
\end{matrix} \right],
\end{equation}
\begin{equation}\label{eq13}
\mathbf{B}=\frac{1}{2}\left[ \begin{matrix}
   \left( r_{1}^{2}-r_{2}^{2} \right)+\left( x_{2}^{2}+y_{2}^{2} \right)-\left( x_{1}^{2}+y_{1}^{2} \right)  \\
   \vdots  \\
   \left( r_{1}^{2}-r_{n}^{2} \right)+\left( x_{n}^{2}+y_{n}^{2} \right)-\left( x_{1}^{2}+y_{1}^{2} \right)  \\
\end{matrix} \right],
\end{equation}
\begin{equation}\label{eq14}
\mathbf{X}={{[x\ y]}^{T}}.
\end{equation}
To estimate the receiver coordinates as $\mathbf{\hat{X}}={{[\hat{x}\ \hat{y}]}^{T}}$, the least square estimation method is employed  to minimize the squared Euclidean distance as $\mathbf{S}=\left\| \mathbf{B-A\hat{X}} \right\|_{2}^{2}$ \cite{18}. Derivation of $\mathbf{S}$ is obtained and by setting $\frac{d\mathbf{S}}{d\mathbf{X}}$ to zero, the following equation is derived
\begin{equation}\label{eq15}
-2{{\mathbf{A}}^{\mathbf{T}}}\mathbf{B}+2{{\mathbf{A}}^{\mathbf{T}}}\mathbf{A\hat{X}}=0.
\end{equation}
A unique solution is finally calculated as
\begin{equation}\label{eq16}
 \mathbf{\hat{X}}={{({{\mathbf{A}}^{\mathbf{T}}}\mathbf{A})}^{-1}}{{\mathbf{A}}^{\mathbf{T}}}\mathbf{B}.
\end{equation}

\section{SIMULATION AND RESULT}
Three typical locations are selected to analyze the effect of multipath reflections. \emph{A} (0, 0,1.2) represents a point at the corner of the room, where the scatterings and reflections are severe; \emph{B} (4, 0, 1.2) represents a point at the edge of the room, right beside the wall, where reflections are medium; and \emph{C} (4, 4, 1.2) represents central point where the effect of multipath reflections becomes weak.
\begin{figure}
\centering
\includegraphics[width = 6cm, height = 6cm]{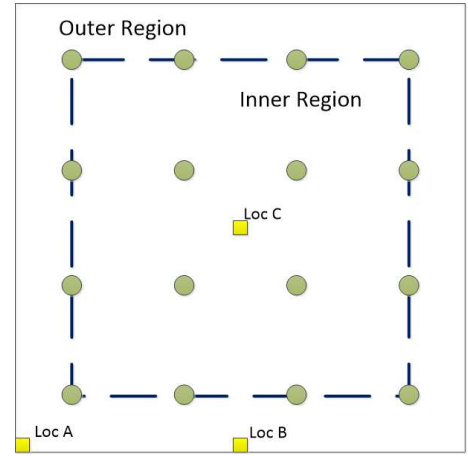}
\caption{Bird’s-eye view of the system model.}
\end{figure}
Fig. 5 shows a bird’s-eye view of the system model. The circles show the locations of all the LED bulbs, and the three selected locations are marked with squares. We assume that a total number of 16 LED bulbs are installed on the ceiling of the room. The inner region is the area within the dashed lines in Fig. 5 while the rest of area is considered as the outer region.
\subsection{Impulse Response Analysis}
Considering the symmetrical property of the room, the impulse responses from the transmitter located at (3, 3, 3. 3) are investigated at the three selected locations. The contribution of the LOS and the first three reflections are shown in Figs. 6-8.

Particularly, Fig. 6 demonstrates the impulse response of each reflection order at Location A. The impulse response amplitude of reflections is comparable to that of LOS incurring large positioning errors.  Fig. 7 shows the impulse response of each reflection order at Location B. The amplitude of the reflections significantly decreases compared to Location A, and thus positioning accuracy is expected to be improved. As can be seen from Fig. 8, the LOS component almost dominates the total impulse response, and the amplitude of the reflections is negligible at Location C. Therefore, the positioning performance is expected to be less affected by multipath reflections.
\begin{figure}
\centering
\includegraphics[width = 7.5cm, height = 6.5cm]{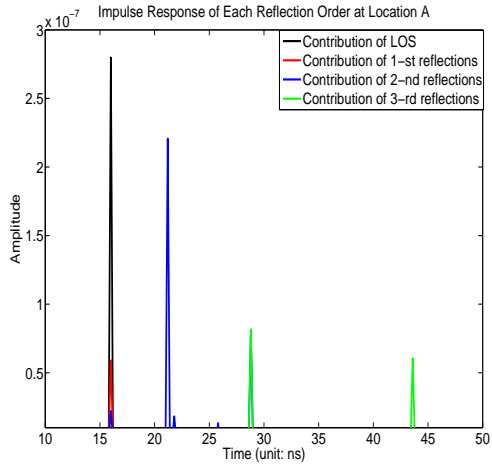}
\caption{Impulse response of each reflection order at Location \emph{A}.}
\end{figure}
\begin{figure}
\centering
\includegraphics[width = 7.5cm, height = 6.5cm]{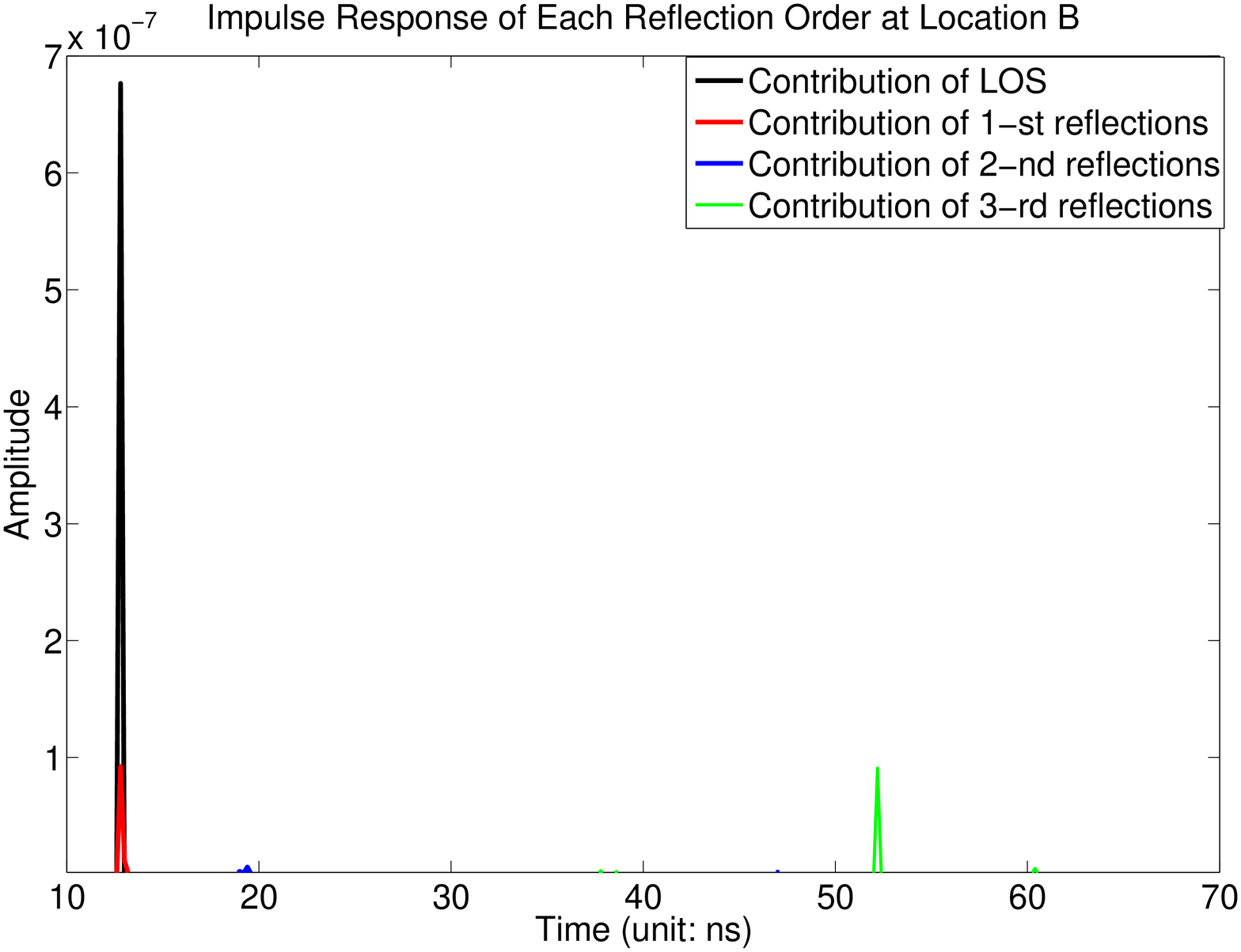}
\caption{Impulse response of each reflection order at Location \emph{B}.}
\end{figure}
\begin{figure}
\centering
\includegraphics[width = 7.5cm, height = 6.5cm]{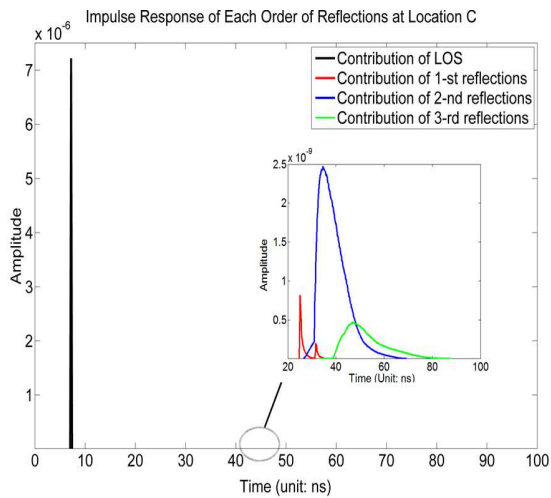}
\caption{Impulse response of each reflection order at Location \emph{C}.}
\end{figure}
\subsection{Power Intensity Distribution Analysis}
As RSS information is applied to estimate the distance between a transmitter and receiver, the received power from each transmitter directly affects positioning performance. In this subsection, we investigate received power from different LED transmitters at each reflection order for the three selected locations. Figs. 9-11 present the highest six received power values for the selected locations inside the room in descending order. The received power of each reflection order at the corner point is shown in Fig. 9. As can be seen clearly, only for the first LED signal, the LOS power value is much greater than that of the reflections. However for the other five LED signals, the reflection components are comparable to the LOS component. The reflection components affect positioning accuracy since only direct power attenuation from the transmitter is considered in distance estimation.
\begin{figure}
\centering
\includegraphics[width = 7.5cm, height = 6.5cm]{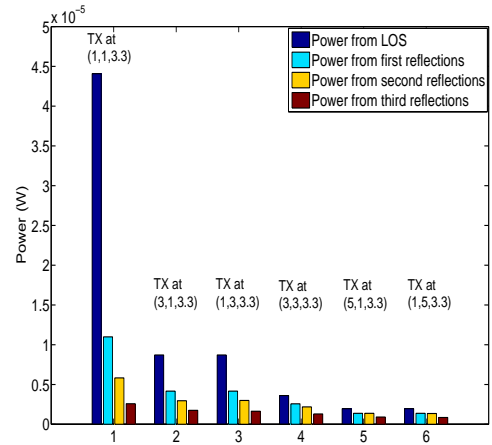}
\caption{Received power from each LED bulb at Location \emph{A}.}
\end{figure}
\begin{figure}
\centering
\includegraphics[width = 7.5cm, height = 6.5cm]{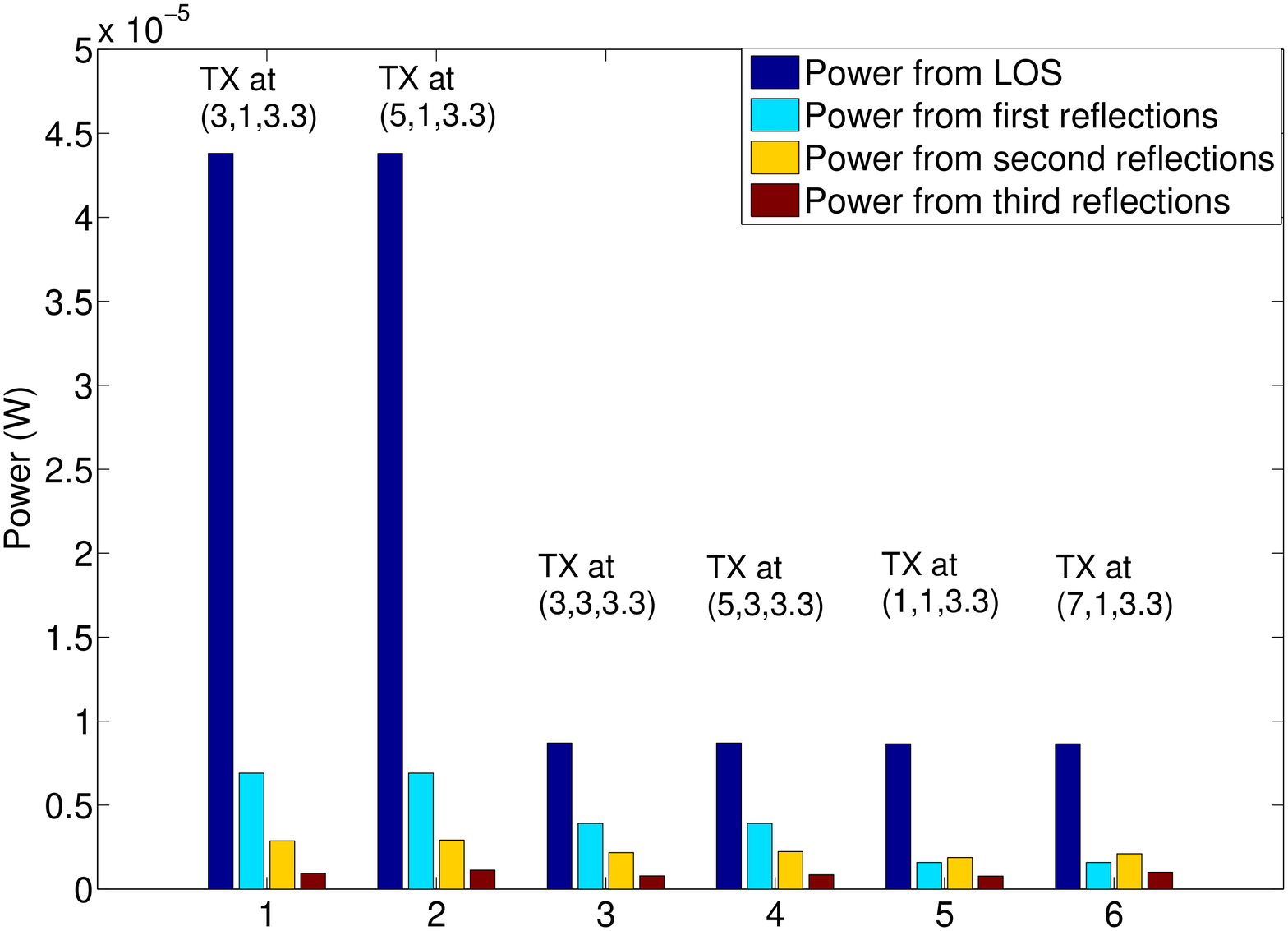}
\caption{Received power from each LED bulb at Location \emph{B}.}
\end{figure}
\begin{figure}
\centering
\includegraphics[width = 7.5cm, height = 6.5cm]{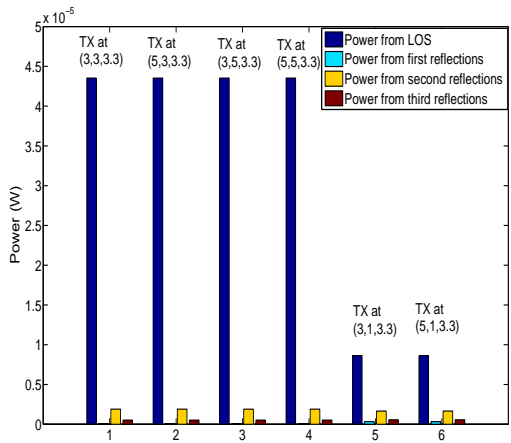}
\caption{Received power from each LED bulb at Location \emph{C}.}
\end{figure}
\subsection{Analysis of Positioning Accuracy}
When distance between a transmitter and receiver is estimated by the received power, the receiver coordinates are finally obtained by the positioning algorithm as stated in Section III.

As a benchmark and in order to show the effect of multipath reflections on the positioning accuracy, positioning error neglecting the reflected power is also calculated and shown in Fig. 12. As can be seen, the positioning error is low all over the room, and only slightly higher in the corner area.
\begin{figure}
\centering
\includegraphics[width = 7.5cm, height = 6.5cm]{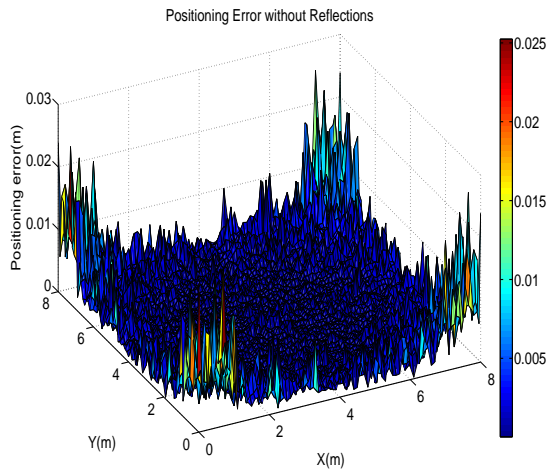}
\caption{Positioning error considering no reflections.}
\end{figure}
Fig. 13, on the other hand, shows the positioning performance considering the multipath reflections. It can be noted that each location of the room is affected by multipath reflections, especially the corner and edge areas. However, the positioning accuracy is satisfactory at the centre of the room as reflections are weak there.
\begin{figure}
\centering
\includegraphics[width = 7.5cm, height = 6.5cm]{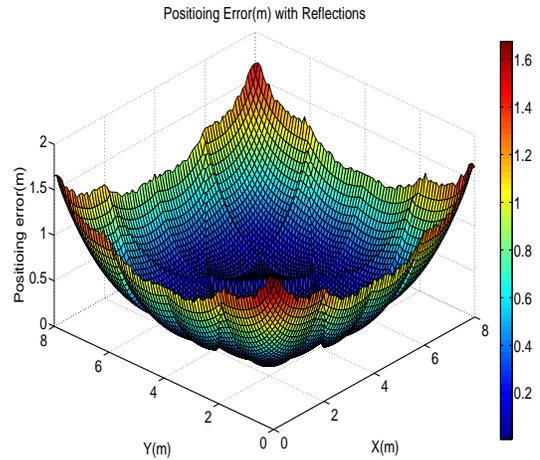}
\caption{Positioning error considering reflections.}
\end{figure}
\begin{figure}
\centering
\includegraphics[width =7.5cm, height = 6.5cm]{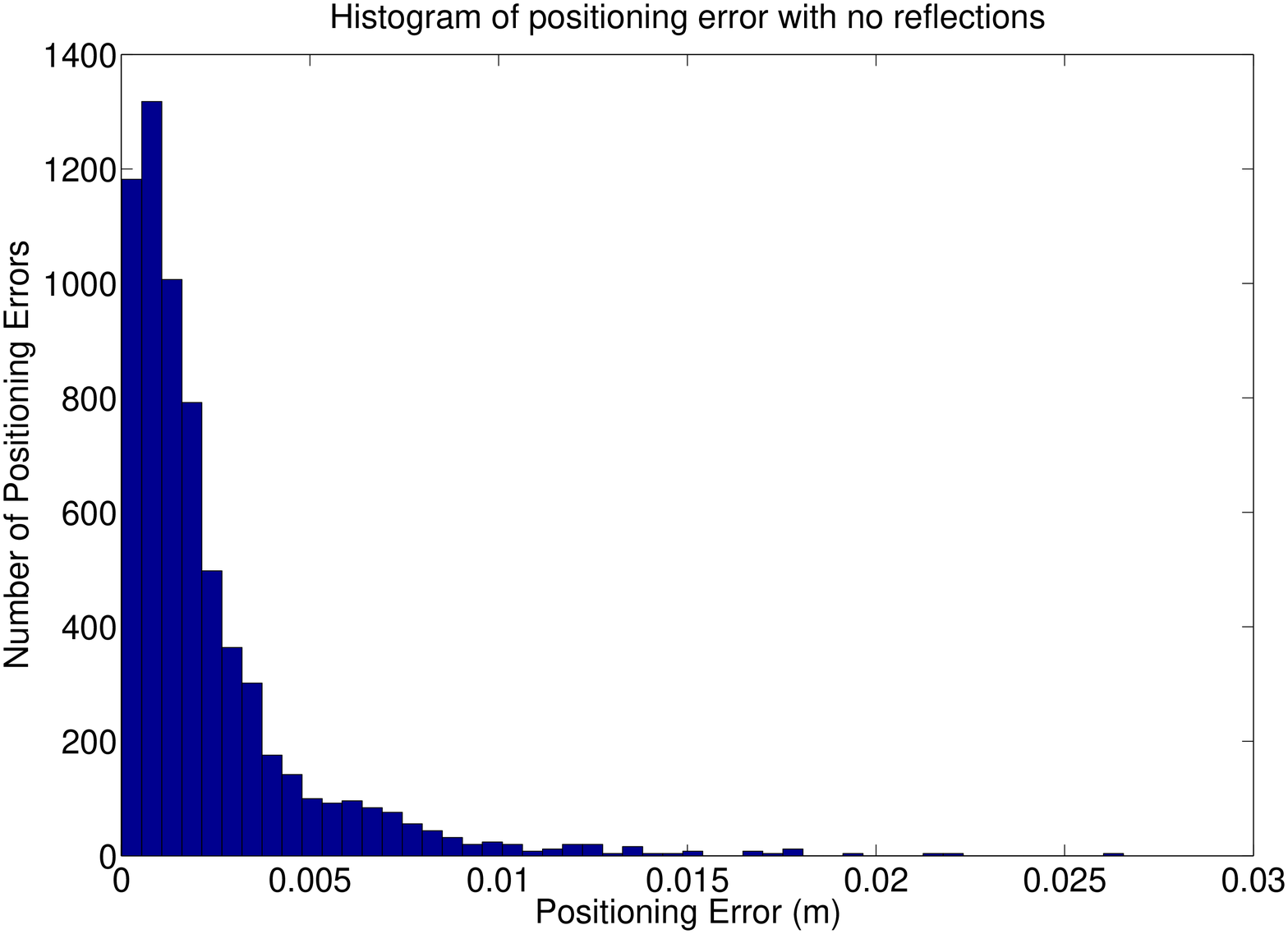}
\caption{Histogram of positioning error considering no reflections.}
\end{figure}
\begin{figure}
\centering
\includegraphics[width = 7.5cm, height = 6.5cm]{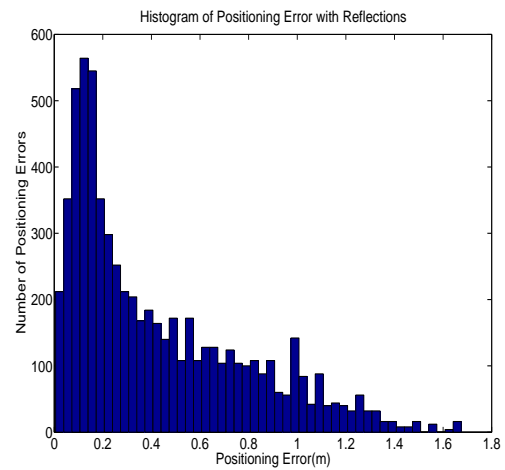}
\caption{Histogram of positioning error considering reflections.}
\end{figure}
Fig. 14 and Fig. 15 present the histograms of positioning errors when neglecting and considering multipath reflections, respectively. When no reflections are considered, the errors only come from the thermal noise and shot noise which are small \cite{19}. In this case, most of the errors are within 0.005 m which is an ideal scenario. However, reflections cannot be practically avoided, and thus they are a major concern in the positioning system impairing dramatically the system performance as shown in Fig. 15. In this case, most of the positioning errors are below 1 m while at some locations, the error climbs up to 1.7 m.
\begin{table}
\begin{center}
\begin{quote}
\caption{Positioning error neglecting/considering reflections}
\label{table3}
\end{quote}
\begin{tabular}{|c|c|c|}
  \hline
  \textbf{RMS error} &\textbf{Reflections}   &\textbf{Reflections}\\
  &\textbf{neglected (m)} & \textbf{included (m)}\\ \hline
  Loc. \emph{A}  &0.0098 &1.6544 \\ \hline
  Loc. \emph{B}  &0.0019 &0.9966 \\ \hline
  Loc. \emph{C}  &0.0012 &0.1674 \\ \hline
  Outer region (RMS)  &0.0059 &0.8173 \\
  & &\\\hline
  Inner region (RMS)  &0.0016 &0.2024 \\
  & & \\\hline
  Entire room (RMS)  &0.0040 &0.5589 \\
   & & \\\hline
  \end{tabular}
\end{center}
\end{table}

Table III compares the positioning error quantitatively when neglecting and considering the reflections. At Location \emph{A}, the root mean square (RMS) error is 1.6544 m since the reflections are strong there. The effect of multipath reflections is medium at Location B while the positioning performance is the best at Location C. For the outer region, the RMS error is 0.8173 m due to severe reflections while the RMS error of the inner region is 0.2024 m. The RMS error for the entire room is 0.5589 m while it is only 0.0040 m when no reflections are considered.
\section{Discussion and Calibration Approaches}
As shown in Section IV, multipath reflections considerably affect the positioning accuracy, especially on the outer region. To improve system performance, three calibration approaches are proposed in this section.
\subsection{Nonlinear estimation}
Previous calculation is based on the linear least square estimation method as stated in Section III. Coordinates in (\ref{eq9}) are obtained by looking for the optimized solution minimizing $\mathbf{S}$. However, the mathematical deduction from (\ref{eq8}) to (\ref{eq9}) is not reversible. In other words, the optimized solution for (\ref{eq9}) is not perfectly suitable for (\ref{eq8}). Therefore, the linear least square estimation may introduce an error in positioning algorithm.
Nonlinear least square estimation is proposed in this section as a calibration approach. Instead of finding the optimized least square solution for (\ref{eq9}), solution for (\ref{eq8}) is directly estimated by finding $\mathbf{\tilde{X}}=\left[ \tilde{x},\tilde{y} \right]$ that minimizes
\begin{equation}\label{eq15}
\mathbf{\tilde{S}}={{\sum\limits_{i}{(F(x,y;{{x}_{i}},{{y}_{i}})-{{r}_{i}})}}^{2}}    ,
\end{equation}
where $F(x,y;{{x}_{i}},{{y}_{i}})=\sqrt{{{(x-{{x}_{i}})}^{2}}+{{(y-{{y}_{i}})}^{2}}}$. Employing the trust region reflective algorithm, an iterative process is applied to estimate $\mathbf{\tilde{X}}$ \cite{20}. Briefly, this algorithm works by first initializing an estimate as ${{\mathbf{\tilde{X}}}_{0}}$, and then calculate the corresponding ${{\mathbf{\tilde{S}}}_{0}}$. Second, several points surrounding ${{\mathbf{\tilde{X}}}_{0}}$ are substituted in (\ref{eq15}), and the one that minimizes ${{\mathbf{\tilde{S}}}_{1}}$ is selected as ${{\mathbf{\tilde{X}}}_{1}}$. After several iterative steps, receiver coordinates $\mathbf{\tilde{X}}$ will finally be obtained when $\mathbf{\tilde{S}}$ converges. In our approach, the value estimated with linear least square estimation in (\ref{eq14}) is selected as the initial value.
\begin{figure}
\centering
\includegraphics[width = 7.5cm, height = 6.5cm]{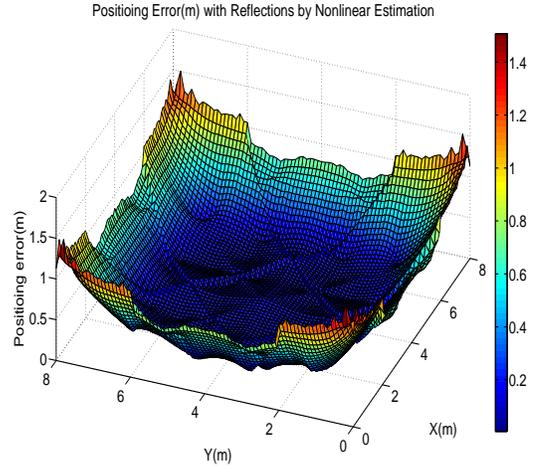}
\caption{Positioning error with nonlinear estimation.}
\end{figure}
\begin{figure}
\centering
\includegraphics[width = 7.5cm, height = 6.5cm]{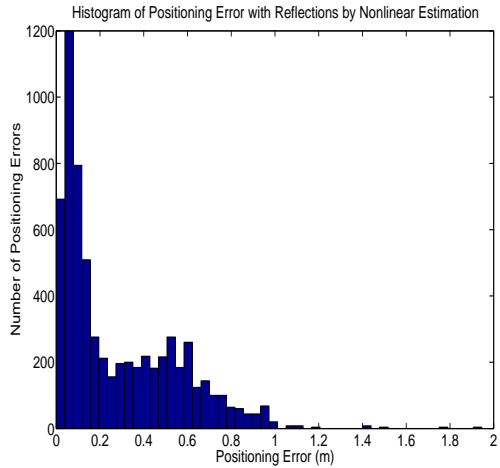}
\caption{Histogram of positioning error with nonlinear estimation.}
\end{figure}
Fig. 16 demonstrates the positioning error distribution with the nonlinear estimation approach. The histogram of the positioning errors is shown in Fig. 17. In this case, most of the errors are within 0.8 m, and only a few of them are over 1 m. The worst positioning error is just around 1.5 m.

Table IV compares the performance of the linear and non-linear estimation methods. As can be seen, the nonlinear estimation outperforms its linear estimation counterpart. Particularly, for the outer region where the reflections are severe, the nonlinear estimation reduces the RMS error to 0.6871 m. The RMS error also decreases to 0.1401 m and 0.4642 for the inner region and the entire room, respectively.
\begin{table}
\begin{center}
\begin{quote}
\caption{RMS error with linear/nonlinear estimation.}
\label{table4}
\end{quote}
\begin{tabular}{|c|c|c|c|}
  \hline
  \textbf{RMS error}&\textbf{Outer}&\textbf{Inner}&\textbf{Entire} \\
  &\textbf{Region(m)}&\textbf{Region(m)}&\textbf{Room (m)} \\ \hline
  Linear   &0.8173 &0.2024 &0.5589 \\
  estimation & & &\\ \hline
  Nonlinear   &0.6871 &0.1401 &0.4642 \\
  estimation & & &\\ \hline
  \end{tabular}
\end{center}
\end{table}
\subsection{Selection of LED Signals}
The received power decreases when distance between a transmitter and receiver increases. As shown in Fig. 5-7, the reflections contribute more to the total received power when the signal is from a farther LED transmitter and brings a larger error in the distance estimation.

In this subsection, a signal selection approach is proposed, i.e., the receiver only selects strong signals for the coordinate estimation. In the numerical analysis, the six, five and four strongest LED signals are selected, and the results are shown in Table V. By removing the signals affected considerably by multipath reflections, the positioning accuracy is improved. The total RMS error decreases to 0.4046 m, 0.3527 m and 0.3240 m respectively for the cases when the six, five and four strongest LED signals are selected. Specially, when the four strongest LED signals are selected, the RMS error is 0.4828 m for the outer region and 0.0849 m for the inner region. As can be noted, the outer region is improved more than the inner region.
\begin{table}
\begin{center}
\begin{quote}
\caption{RMS error with transmitter selection approach.}
\label{table5}
\end{quote}
\begin{tabular}{|c|c|c|c|}
  \hline
  \textbf{RMS error}&\textbf{Outer}&\textbf{Inner}&\textbf{Entire} \\
     &\textbf{Region(m)}&\textbf{Region(m)}&\textbf{Room (m)} \\ \hline
  6 LEDs (Linear   &0.6760 &0.1640 &0.4616 \\
  estimation) & & &\\ \hline
  6 LEDs (Nonlinear   &0.6016 &0.1112 &0.4046 \\
  estimation) & & &\\ \hline
  5 LEDs (Linear   &0.5472 &0.1251 &0.3722 \\
  estimation) & & &\\ \hline
  5 LEDs (Nonlinear   &0.5272 &0.0851 &0.3527 \\
  estimation) & & &\\ \hline
  4 LEDs (Linear   &0.4838 &0.0933 &0.3259 \\
  estimation) & & &\\ \hline
  4 LEDs (Nonlinear   &0.4828 &0.0849 &0.3240 \\
  estimation) & & &\\ \hline
  \end{tabular}
\end{center}
\end{table}
For the sake of presentation, only the best results are shown in Figs. 18 and 19. Fig. 18 shows the positioning error distribution when the four strongest LED signals are selected for distance calculation and the nonlinear estimation is applied to obtain receiver coordinates. Fig. 19 presents the corresponding histogram of positioning errors. It can be seen from Fig. 19 that many of the locations have positioning error which is less than 0.4 m, and only a few locations have positioning error that is larger than 0.8 m.
\begin{figure}
\centering
\includegraphics[width = 7.5cm, height = 6.5cm]{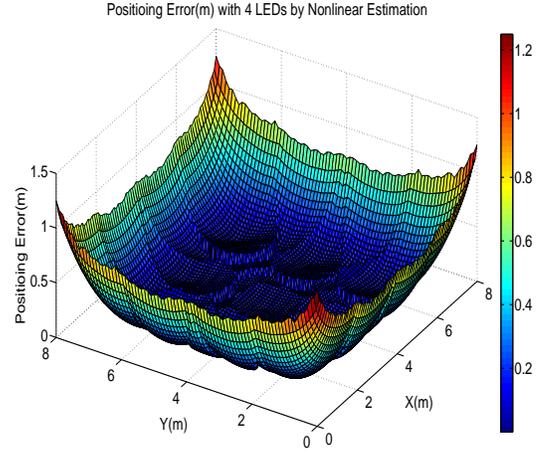}
\caption{Positioning error (4 selected LED signals and nonlinear estimation).}
\end{figure}
\begin{figure}
\centering
\includegraphics[width = 7.5cm, height = 6.5cm]{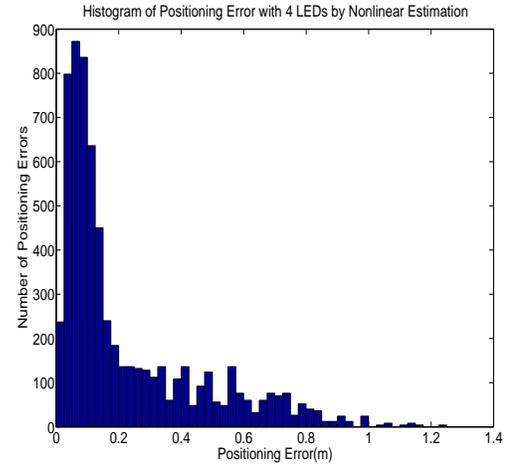}
\caption{Histogram of positioning error (4 selected LED signals and nonlinear estimation).}
\end{figure}
\subsection{Deceasing the Distance between LED Bulbs}
When LED bulbs are installed in a dense layout (i.e., the distance between LED bulbs is reduced, and a greater number of LED bulbs are used), the light intensity distribution becomes more uniform for the entire room, and therefore, the positioning accuracy is improved.

Table VI shows the RMS error where distance between the LED bulbs decreases to 1.5 m, and 25 LED bulbs are installed in total. With no LED signal selection, the entire RMS error is 0.3121 for the nonlinear estimation. The RMS error decreases to 0.2922 m and 0.2699 m when six and five LED signals are selected, respectively. However, when the four strongest LED signals are selected, the receiver coordinates cannot be obtained by the linear estimation in some locations. The shorter distance between the LED bulbs increases the probability that three of the selected LED bulbs are in the same row. In this scenario, there is a singularity in Matrix $\mathbf{A}$, and the coordinates estimation fails. If this case happens, the average of the detected LED bulb coordinates will be applied as the initial value instead of the linear estimated value. As this initial value is not as accurate as the linear estimated one, the positioning error increases as shown in Table VI.
\begin{table}
\begin{center}
\begin{quote}
\caption{RMS error with 1.5 m distance between the LED bulbs.}
\label{table6}
\end{quote}
\begin{tabular}{|c|c|c|c|}
  \hline
  \textbf{RMS error}&\textbf{Outer}&\textbf{Inner}&\textbf{Entire} \\
   &\textbf{Region(m)}&\textbf{Region(m)}&\textbf{Room (m)} \\ \hline
  All LEDs (Linear   &0.6049 & 0.1093 & 0.4064 \\		
  estimation) & & &\\ \hline
  All LEDs (Nonlinear   &0.5310 &0.0861 &0.3553 \\		
  estimation) & & &\\ \hline
  6 LEDs (Linear   &0.4400 &0.0916 &0.2976 \\		
  estimation) & & &\\ \hline
  6 LEDs (Nonlinear   &0.4362 &0.0729 &0.2922 \\	 	
  estimation) & & &\\ \hline
  5 LEDs (Linear   &0.4106 &0.0691 &0.2751 \\		
  estimation) & & &\\ \hline
  5 LEDs (Nonlinear   &0.4014 &0.0742 &0.2699 \\		
  estimation) & & &\\ \hline
  4 LEDs (Linear   &NA &0.0571 &NA \\	
  estimation) & & &\\ \hline
  4 LEDs (Nonlinear   &0.4251 &0.0729 &0.2850 \\ 		
  estimation) & & &\\ \hline
  \end{tabular}
\end{center}
\end{table}
\begin{figure}
\centering
\includegraphics[width = 7.5cm, height = 6.5cm]{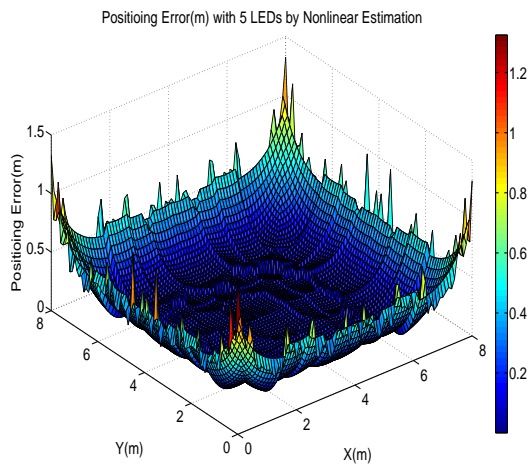}
\caption{Positioning error with 1.5 m distance between LED Blubs (5 selected LED Signals and Nonlinear Estimation).}
\end{figure}
\begin{figure}
\centering
\includegraphics[width = 7.5cm, height = 6.5cm]{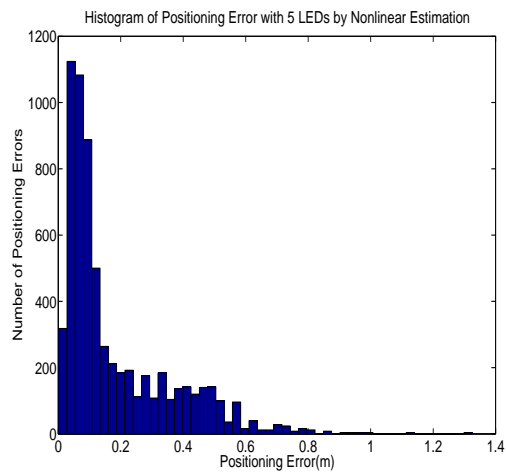}
\caption{Histogram of Positioning Error with 1.5 m Distance between LED Blubs (5 selected LED Signals and Nonlinear Estimation).}
\end{figure}
Figs. 20 and 21 show the positioning performance for the best scenario, i.e., 5 LED signals selected for the nonlinear estimation. As can be seen from Fig. 20, for the most of the inner area, the positioning performance is satisfactory while at the edges and corners of the room, there are some locations with large positioning errors. From Fig. 21, it can be observed that most of the positioning errors are within the range of 0.5 m demonstrating improvement in the positioning performance and usefulness of the proposed method.
\section{Conclusions}
In this paper, an indoor visible light positioning system taking account of multipath reflections has been investigated for a typical room where the impulse response is obtained employing CDMMC approach. In particular, the impulse response of each order has been calculated for three locations representing the corner, edge and center points. The received signal power distribution at three locations with different degrees of multipath reflections has been analyzed. Furthermore, the positioning error in the entire room has been calculated and compared with the previous works where reflections are not taken into account. Comparison has shown that multipath reflections considerably decrease the positioning accuracy, especially for the outer region. Three calibration approaches have been proposed to improve the positioning performance. Particularly, we have shown that by employing nonlinear estimation, the positioning error is decreased. A selection on the LED signals has been also proposed to remove the signals that are severely affected by multipath reflections. Finally, we have shown that a dense layout of LED bulbs results in a more uniform light intensity distribution, and therefore a higher positioning accuracy can be achieved.
\section*{Acknowledgement}
The authors would like to thank the National Science Foundation (NSF) ECCS directorate for their support of this work under Award \# 1201636, as well as Award \# 1160924, on the NSF "Center on Optical Wireless Applications (COWA – http://cowa.psu.edu)"
\balance
\bibliographystyle{IEEEtran}
\bibliography{ref}

\end{document}